\documentclass[prl,twocolumn,amsmath,amssymb,aps,showpacs]{revtex4-1}
\usepackage{graphicx}
\usepackage{color}

\newcommand{\ket}[1]{ {\vert#1\rangle}}
\newcommand{\bra}[1]{ {\langle#1\vert}}
\newcommand{\braket}[2]{ {\langle#1\vert#2\rangle}}

\begin{document}

\title{Variational Numerical Renormalization Group: 
Bridging the gap between NRG and Density Matrix Renormalization Group}
\author{Iztok Pi\v{z}orn}
\author{Frank Verstraete}
\affiliation{University of Vienna, Faculty of Physics, Boltzmanngasse
5, A-1090 Wien (Austria)}
\date{May 11, 2011}

\pacs{05.10.Cc, 72.10.Fk, 72.15.Qm, 02.70.-c, 03.67.-a,71.27.+a}

\begin{abstract}
The numerical renormalization group (NRG) is rephrased as a variational method with the cost function given by the sum of all the energies of the effective low-energy Hamiltonian. This allows to systematically improve the spectrum obtained by NRG through sweeping.
The ensuing algorithm has a lot of similarities to the density matrix renormalization group (DMRG) when targeting many states, and this synergy of NRG and DMRG combines the best of both worlds and extends their applicability. We illustrate this approach with simulations of a quantum spin chain and a single impurity Anderson model (SIAM) where the accuracy of the effective eigenstates is greatly enhanced as compared to the NRG, especially in the transition to the continuum limit.
\end{abstract}

\maketitle


The density matrix renormalization group \cite{white}  (DMRG),
devised to improve on the Wilson's numerical renormalization group (NRG) \cite{wilson},
has become the method of choice to simulate 
one-dimensional quantum many-body systems at zero temperature and has found a large number 
of applications in the fields of the condensed matter physics, quantum chemistry and quantum information theory where it turned out that DMRG is essentially equivalent to simulating quantum systems in terms of matrix product states (MPS) \cite{mpsbegin}.
In all these methods, a quantum many-body state is represented by associating matrices to local configurations at sites and the coefficients in the expansion over the configuration states are given as products of the corresponding matrices.
The DMRG can be used to calculate not only the ground state but also the excited states through the concept of \emph{targeting} where these matrices are chosen such that they well represent many states at the same time. In the NRG, on the other hand, the low energy states of a system are expressed in terms of the low energy states of a smaller system
which is a suitable description of impurity systems with energy scale separation. In this context, the NRG gives remarkably good results and has retained the position as a widely used impurity solver \cite{bullareview} whereas the DMRG is rarely used to calculate the excited states \cite{schneider} except for the spectral gap.
One of the reasons is that the NRG is less costly as it provides $D$ excited states at the cost of $O(D^3)$ as compared to DMRG with targeting with the cost $O(D^4)$.
Presently, the DMRG is mostly used in its time dependent form (see e.g.~\cite{dmrgreview}) which is also true for impurity systems \cite{ddmrg,nishimoto,franksiam,karski,raas} where also other density-matrix related concepts are used \cite{hofstetter, andersschiller,saberi}.
Still, DMRG's remarkable ability to calculate and optimize many excited states has not been used in this context.

Even as an impurity solver, the NRG has certain limitations: 
the hopping terms should fall off sufficiently fast which means lower resolution of the spectral densities at higher energies e.g. Hubbard bands. Otherwise it becomes inaccurate for longer chains which is seen as a violation of the Friedel sum rule \cite{zitko}. It also becomes highly expensive for 
many-band impurity problems whereas an additional self-consistency constraint in the 
context of dynamical mean-field theory, see e.g. \cite{dmftreview}, calls for 
more accurate impurity solvers.
What makes the NRG really different from the DMRG is that it does not provide any feedback mechanism to optimize the matrices by sweeping along the chain and a method along this line has already been used with quantum fields \cite{konik}. In this Letter, we identify the cost function in the NRG algorithm and introduce a feedback mechanism by which the states can be optimized in a variational way under the original NRG cost function. We relate the NRG and the DMRG by identifying the common cost function in both approaches show that the proposed scheme improves the results of both methods while retains the lower, NRG-like, scaling of computational costs.


\textit{Numerical renormalization group.} In the context of the NRG, the lowest energy eigenstates 
of a quantum many-body system on $n$ sites, 
$\hat{H} \ket{\psi_\alpha} = E_{\alpha} \ket{\psi_\alpha}$,
are approximated by orthonormal
states $\mathcal{S}^{[n]} \equiv \{ \psi^{[n]}_1, \ldots, \psi^{[n]}_{D_{n}}\}$ defining an effective  low energy Hamiltonian
\begin{equation}
\hat{H}_{\rm eff}^{[n]} = \sum_{\alpha=1}^{D_n} E^{[n]}_\alpha \ket{\psi^{[n]}_\alpha} \bra{\psi^{[n]}_\alpha} 
\label{eq:Heff}
\end{equation}
with effective energies $E^{[n]}_\alpha = \bra{\psi^{[n]}_\alpha } \hat{H} \ket{\psi^{[n]}_\alpha}$.
The set of effective states $\mathcal{S}^{[n]}$ is 
obtained recursively from $\mathcal{S}^{[n-1]}$, the effective description of a system on $n-1$ sites, extended by an extra site of a local dimension $d_n$, resulting in an extended Hamiltonian 
$\mathbf{H}_{\rm ext}^{[n]} \in \mathbb{C}^{ (D_{n-1} d_n) \times (D_{n-1} d_n) }$
that, in order to prevent exponential growth, has to be projected to a subspace of a smaller dimension $D_n$.
In the NRG, this subspace is spanned by the lowest energy eigenvectors as 
$\mathbf{H}_{\rm eff}^{[n]} = \mathbf{U}^H \mathbf{H}_{\rm ext}^{[n]} \mathbf{U}$ where 
$\mathbf{H}_{\rm eff}^{[n]} \in \mathbb{C}^{ D_n \times D_n }$ is a new effective Hamiltonian and  $\mathbf{U} \in \mathbb{C}^{ (D_{n-1} d_n) \times D_n}$
is an isometry matrix $\mathbf{U}^H \mathbf{U} = \mathbf{1}$.
The set of effective states $\{ \psi^{[n]}_\alpha \}$ is given as a MPS with an external $\alpha$-index \cite{franksiam} 
\begin{equation}
\ket{\psi_\alpha^{[n]}} = \sum_{\{s_j\}} \mathbf{e}_0 \cdot \mathbf{A}^{[1] s_1} \mathbf{A}^{[2] s_2} \cdots \mathbf{A}^{[n] s_n} \mathbf{e}_\alpha \ket{s_1,\ldots s_n}
\label{eq:Psin}
\end{equation}
with the right boundary vector $\mathbf{e}_\alpha$ enumerating basis states. The set $\mathcal{S}^{[n]}$ is orthonormal due to the isometry constraint for matrices $\mathbf{A}^{[j]}$ defined as 
$[ \mathbf{A}^{[j]} ]_{(l s) r} \equiv A^{[j] s}_{l r}$.

The NRG projection is equivalent to requiring that the \emph{sum} of the new effective energies 
$E_\alpha^{[n]}$ is minimal which, according to~(\ref{eq:Heff}), corresponds to the trace of $H_{\rm eff}^{[n]}$ and leads to a cost function
\begin{equation}
f(\mathbf{A}) = {\rm tr} ( \mathbf{A}^H \mathbf{H}_{\rm ext}^{[n]} \mathbf{A} )
\quad\textrm{where}\quad \mathbf{A}^H \mathbf{A} = \mathbf{1}
\label{eq:cost1}
\end{equation}
which is minimized exactly by the NRG isometry $\mathbf{U}$. This cost function is different from other DMRG-based approaches (e.g.~\cite{franksiam,saberi})  where the cost is determined from the ground state alone.


\textit{Variational optimization.} 
The identification of the NRG cost function allows us to optimize not only 
the tensor associated with the last site in the NRG-MPS~(\ref{eq:Psin}) but an arbitrary tensor in the chain.
\begin{figure}
\centering
\includegraphics[width=0.8\columnwidth]{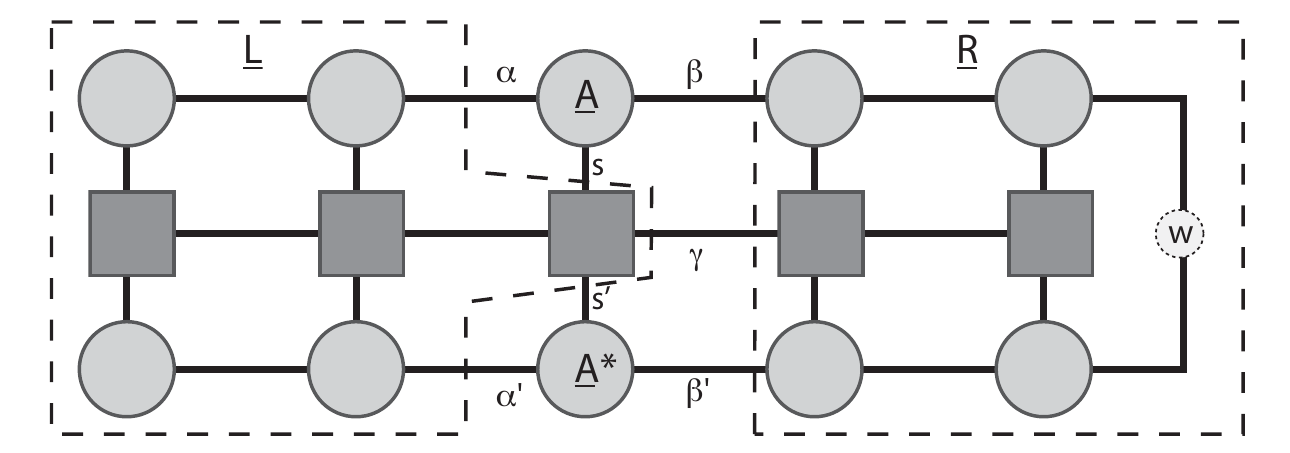}
\caption{Tensor network representation of the (weighted) cost function~(\ref{eq:gencost}) in the optimization of a tensor $\underline{A}^{[j]}$ (circles) under the Hamiltonian matrix product operator (squares).}
\label{fig:trH}
\end{figure}
The cost  $f(\mathbf{A}^{[j]}) = \sum_{\alpha} \bra{\psi^{[n]}_\alpha} \hat{H} \ket{\psi^{[n]}_\alpha}$ can easily be understood as a tensor network depicted on Fig.~\ref{fig:trH} where the $\alpha$-index enumerating states $\ket{\Psi^{[n]}_{\alpha}}$ is contracted with the $\alpha$-index in the conjugate MPS $\bra{\Psi^{[n]}_\alpha}$.
The cost can thus be minimized by varying an arbitrary isometry $A^{[j]}$ in the NRG-MPS~(\ref{eq:Psin}) which we will illustrate by expressing the Hamiltonian as a matrix product operator (any 1D operator can be written as a MPO)
\(
\hat{H} = \sum_{\{\underline{s}_j\}} \mathbf{e}_0 \cdot \mathbf{H}^{[1] \underline{s}_1 } \cdots \mathbf{H}^{[n] \underline{s}_n} \mathbf{e}_0\, 
\ket{s_1',\ldots,s_n'}\bra{s_1,\ldots,s_n}
 \)
where tensors $\underline{H}^{[j]}$ are represented by squares on Fig.~\ref{fig:trH}.
An arbitrary isometry $\mathbf{A}^{[j]}$ is extracted out of the cost function~(\ref{eq:cost1})  which now takes a form
\begin{equation}
f(\mathbf{A}) = \sum_\gamma {\rm tr}( \mathbf{L}_\gamma^{[j]} \mathbf{A} \mathbf{R}_\gamma^{[j]\, T} \mathbf{A}^{H} )
\quad\textrm{for}\quad \mathbf{A}^{H} \mathbf{A} = \mathbf{1}
\label{eq:gencost}
\end{equation}
where $\mathbf{L}^{[j]}$ and $\mathbf{R}^{[j]}$
correspond to the contractions of the tensor networks on the left and right side of the chosen site $j$, respectively (Fig.~\ref{fig:trH}), explicitly defined as
$[\mathbf{L}_\gamma^{[j]}]_{(\alpha' s')(\alpha s)} = L_{\alpha s \gamma s' \alpha'}$ and $[\mathbf{R}_\gamma^{[j]}]_{\beta' \beta} = R_{\beta\gamma\beta'}$.
Furthermore, a modified weighted cost function may be considered where the effective states are weighted according the their importance, e.g. by a Boltzmann factor $w_\alpha = e^{-\beta E^{[n]}_{\alpha}}$ (see Fig.~\ref{fig:trH}). 
The cost~(\ref{eq:gencost}) is minimized using the conjugate gradient method with a unitary constraint \cite{abrudanCG} in a variational way (i.e. the sum of energies can only decrease) and the computational costs scale as $O(D^3)$, same as in the original NRG.
The number of optimization steps depends on the quality of the initial state and the desired accuracy.

\textit{Density matrix renormalization group.}
A special property of the NRG-MPS states is a two-fold nature of the external $\alpha$-index which acts both as an enumerating index as well as a virtual bond in extending the system for a site. 
For a fixed system size, the external index can be associated with any site in the chain and moved along the chain by means of the singular value decomposition (SVD). 
Interestingly, the concept of a moveable external index is intrinsic to the finite size DMRG algorithm \emph{with targeting} \cite{white}.
The basic concept of the finite size DMRG is as follows: split the system as  $\{1,\ldots,j-1\}, \{j\}, \{j+1\}, \{j+2,\ldots,n\}$, find the optimal tensor at site $j$ and move to the next site $j + 1$.
The way one finds the optimal tensor at site $j$ is by considering the reduced density matrix (DM) for $\{1,\ldots,j\}$, obtained by tracing the DM of the complete system (``universe'') over the environment $\{j+1,\ldots,n\}$ whereas the universe can be either in a pure state (ground state) or in a mixed state of lowest energy states as is the case with the targeting.
Formally, the eigenstates of the universe can be written as $\ket{\Psi_\alpha} = \sum  G_{(l,s_j,s_{j+1},r), \alpha} \ket{\varphi_l^{1,\ldots,j-1}}\ket{s_j}\ket{s_{j+1}}\ket{\varphi_r^{j+2,\ldots,n}}$ where the index $\alpha$ enumerates the states and the optimal tensor $\underline{A}^{[j]}$ is obtained by the SVD as $G_{(l s_j s_{j+1} r) \alpha} = \sum_c A^{[j] s_j }_{l c} \sigma_c V_{c s_{j+1} \alpha r}$. Identifying $A^{[j+1] s_{j+1} \alpha}_{l,r} \equiv \sigma_l V_{l s_{j+1} \alpha r}$, we recover a NRG-MPS with the index $\alpha$ associated with the site $j+1$ (instead of site $n$). However, this part is ignored in the DMRG since the tensor at site $j+1$ is obtained in the subsequent step. Therefore, the index $\alpha$ is intrinsic to but hidden in the DMRG and is carried back and forth along the chain. 
In the DMRG, only the central sites are optimized (the boundary parts are treated exactly).
We can however sweep to the very end of the chain and end up exactly with the NRG-MPS~(\ref{eq:Psin}) where the last tensor $\underline{A}^{[n]}$ carries the $\alpha$ index which now allows us to enlarge the chain for an extra site by a NRG step. Looking back, we realize that the optimal tensor $G_{(l s_j s_{j+1} r)\alpha}$ is an isometry matrix minimizing the same cost function~(\ref{eq:cost1}) as in the NRG context, just that it represents an arbitrary pair of neighboring sites instead of the last site only. Hence, the DMRG minimizes the same cost function as the NRG.

We also realize that the truncation in the SVD decreases the accuracy of states and it is not guaranteed that the optimization at the next site will recover the same accuracy (in fact the highest accuracy is reached in the center of the chain). Therefore, the set of excited states obtained by the DMRG can be further optimized by means of the variational principles proposed earlier.

\textit{Results.} The variational optimization algorithm for a NRG-MPS set of states is put to the test on two qualitatively distinct models. 
First, we consider a quantum Ising chain in a tilted magnetic field
\begin{equation}
H = \sum_{j=1}^{n-1} \sigma_j^x \sigma_{j+1}^x 
+ \sum_{j=1}^{n}  (h_x \sigma_j^{x} + h_z \sigma_j^z)
\label{eq:tilted}
\end{equation}
which is a simple example of non-integrable quantum spin chains.
\begin{figure}
\centering
\includegraphics[width=0.98\columnwidth]{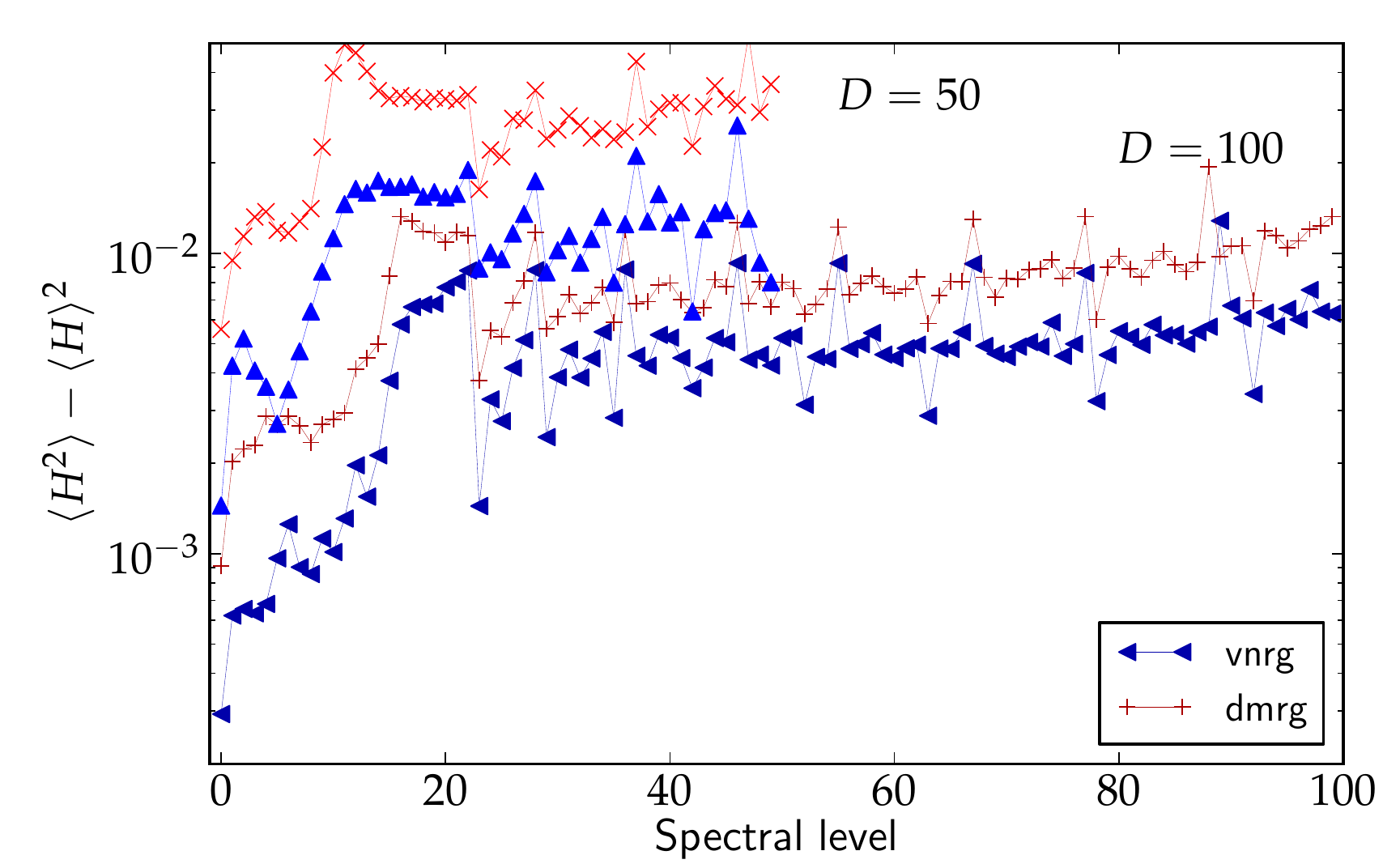}
\caption{Accuracy of the low energy spectral levels for the tilted quantum Ising chain~(\ref{eq:tilted}) on $200$ sites with $h_x=h_z=1$, obtained by the DMRG (red) and optimized by the variational method (blue), for various bond dimensions $D$. }
\label{fig:tilted1}
\end{figure}
The accuracy of the approximate eigenstates of~(\ref{eq:tilted}) is quantified by a measure $\langle H^2 \rangle - \langle H\rangle^2$ which puts a lower bound to the fidelity (see supplementary material.
We consider a chain of $200$ sites, obtain the excited states by means of the DMRG and optimize them using the variational approach. The DMRG provides provides highly accurate results as seen from the Fig.~\ref{fig:tilted1}, leaving little space for improvements. However, the results are not optimal and can be improved by the variational NRG. Higher improvement is observed for smaller bond dimensions $D$ which suggests a possibility of improving the excited states in the regime where sufficiently large $D$ are not reachable due to higher entanglement, such as in two-dimensional or critical (see supp. material) systems.
It is also worth noting that the computational costs of one DMRG sweep with targeting $M$ states scale as $O(n d^3 M D^3)$,  compared to either $O(M D^3 + n d^3 D^3)$ or 
$O(n d^3 D^3)$ for the variational optimization where the $\alpha$-index is associated with an inner or a boundary site, respectively. The pre-factors to $O(M D^3)$ are similar (related to the number of Lanczos steps).

As the second example we consider the Single Impurity Anderson model (SIAM) which, after a logarithmic discretization \cite{bullareview} of the conductance band,
is described by a semi-infinite linear chain where the impurity ($f_{\sigma}$) is coupled to a chain of fermions ($c_{j,\sigma}$) as
\begin{equation}
H_N = \epsilon_f n + U n_{\uparrow} n_{\downarrow} + \!
\sum_\sigma\!\Big( \chi f^\dagger_\sigma c_{0\sigma} + \!\!\sum_{j=0}^{N-1} \! t_j c_{j\sigma}^\dagger c_{j+1\sigma} + \textrm{h.c.} \Big)
\label{eq:SIAM}
\end{equation}
with $n_\sigma \equiv f_\sigma^\dagger f_\sigma$, $n=n_\uparrow + n_\downarrow$, $\chi \equiv \sqrt{\frac{\xi_0}{\pi}}$, and 
$t_j \propto \Lambda^{-j/2}$ (see \cite{bullareview} for details).
The NRG Hamiltonian $H_N$ presents an effective description of the Anderson problem in the energy interval $[\Lambda^{N+1},\Lambda^{N}]$ and only the complete sequence $( H_0, H_1, \ldots )$ gives the complete spectrum. As mentioned in the introduction, the NRG works best for strong scale separation, i.e. $\Lambda \gg 1$ which however take one further away from the continuum limit $\Lambda \to 1$ and yield lower resolution at higher frequencies.
Contrarily, the DMRG does not rely on energy scale separation and thus quickly falling hopping terms but is numerically more costly, scaling as $O(D^4)$ (here $M=D$).
This suggests a great potential in optimizing the \emph{existing} NRG set of states using the variational method proposed in this Letter where the costs only scale as $O(D^3)$, as in the NRG, but, like in the DMRG, no scale separation is required.
Let us first consider the exactly solvable non-interacting SIAM Hamiltonian with $U=0$ with a discretization parameter $\Lambda = 1.7$.
\begin{figure}
\centering
\includegraphics[width=0.98\columnwidth]{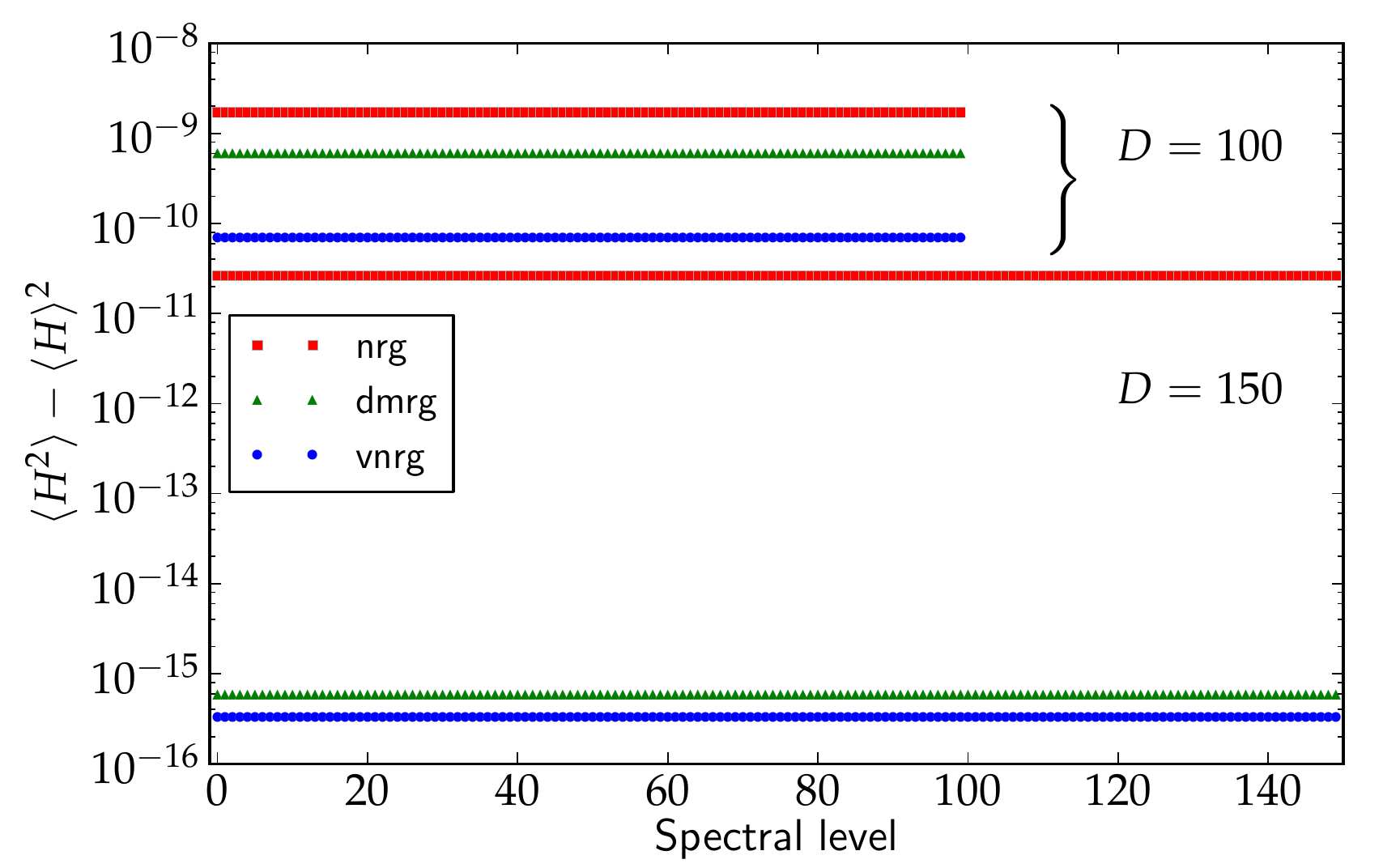}
\caption{Accuracy of the eigenstates for non-interacting SIAM chain~(\ref{eq:SIAM}) on $70$ sites ($N=68$) for NRG, DMRG, and the variational NRG, for $D \in \{100,150\}$. We have set $\Lambda=1.7$, $\xi_0 = 0.1$, $\epsilon_f=-0.1$.}
\label{fig:nisiam}
\end{figure}
We observe from Fig.~\ref{fig:nisiam} that for a fixed bond dimension of a NRG-MPS, the accuracy of eigenstates obtained by the DMRG and the variational optimization (vNRG) is several orders of magnitude better than the NRG results. Qualitatively the same results are obtained for the absolute errors of energies (see supplementary material).

\begin{figure}
\centering
\includegraphics[width=0.98\columnwidth]{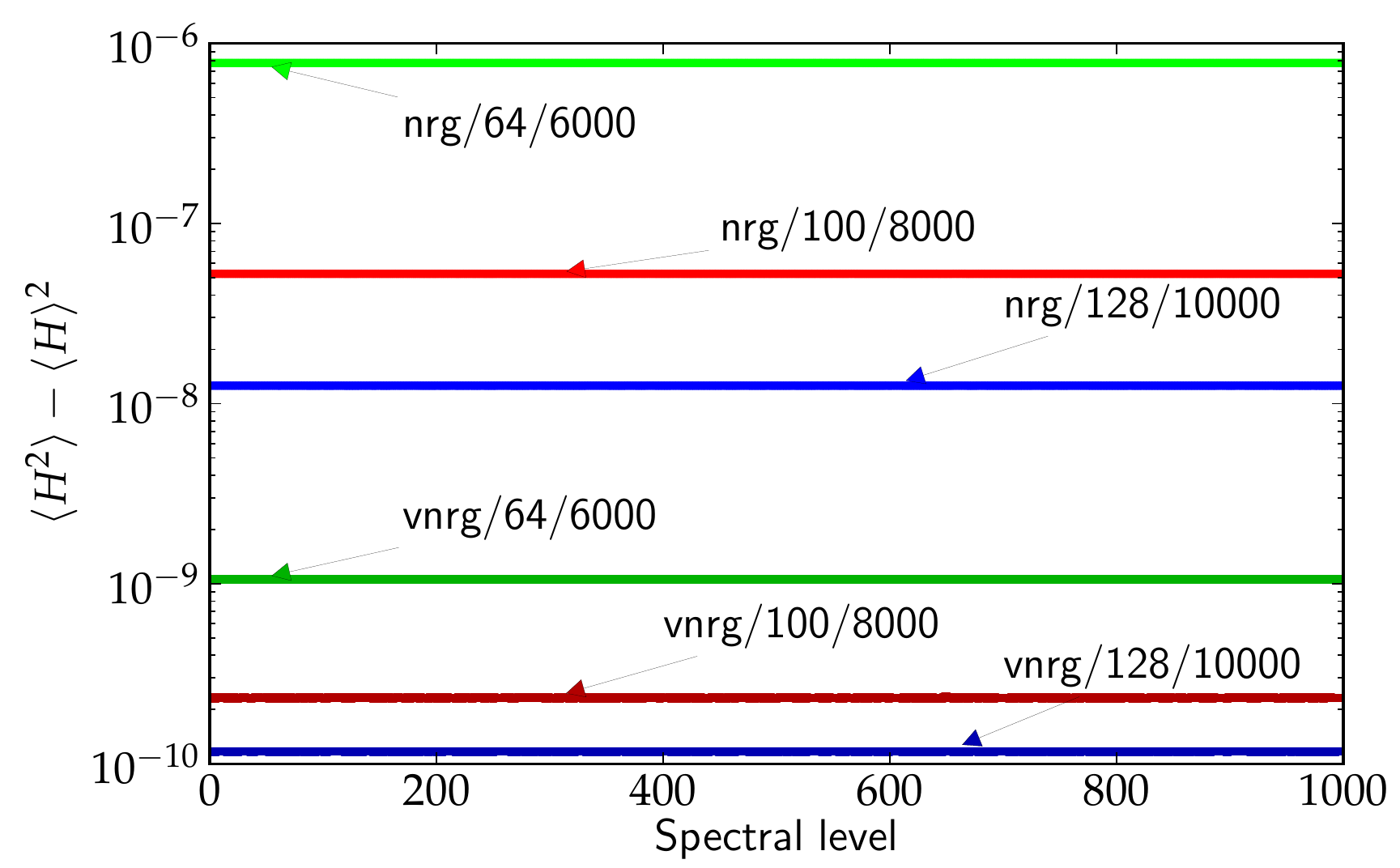}
\caption{
Accuracy of the lowest $1000$ states for the interacting SIAM chain~(\ref{eq:SIAM}) on $40$ sites ($N=38$), for NRG and the variational NRG \emph{with symmetries}. 
Sub-sector bond dimensions are taken $D \in\{64,100,128\}$, truncated to $\{ 6000,8000,10000 \}$ states of lowest energy. 
We have set $\Lambda=2$, $\xi_0 = 0.01$, $\epsilon_f=-0.05$ and $U=0.1$.
}
\label{fig:siam}
\end{figure}

We now consider the interacting SIAM Hamiltonian~(\ref{eq:SIAM}) with $U=-2\epsilon_f = 0.1$ on $40$ sites where, unlike the non-interacting case, we employ the symmetries (total particle number of either spin). We analyze the improvement of the first one thousand NRG states after variational optimization (Fig.~\ref{fig:siam}) in the regime where NRG already produces accurate results due to scale separation ($\Lambda=2$). Due to the symmetries, the bond dimension $D \in \{64,100,128\}$  (which determines the computational complexity) now refers to the maximal number of states in individual subsectors whereas the total number of kept states is denoted by $M \in \{6000,8000,10000\}$.
As shown on Fig.~\ref{fig:siam}, the accuracy of eigenstates is greatly enhanced by the optimization.
While the cost of an optimization step is comparable to the cost of a NRG step, a smaller bond dimension is needed to represent states using the vNRG as compared to the NRG. This effect is more pronounced for small values of $\Lambda$ while for larger values of $\Lambda$ the NRG already gives highly accurate results, only requiring little improvement. 
The results are consistent also for the fidelity of excited states and the absolute errors of energies (see supplementary material).
We stress that the logarithmic discretization is not needed for the variational scheme proposed in this Letter.

\textit{Conclusion.}
We have proposed a variational method to simulate or optimize the effective low energy description of one dimensional quantum systems, introducing a feedback mechanism to the NRG and identifying the cost function. Furthermore, we have expressed the DMRG method with targeting as a NRG method with a moveable external index enumerating states and shown that the DMRG results can be further improved by the variational optimization. We have tested the methods on the quantum Ising chain in a tilted magnetic field and the Single Impurity Anderson model where we observed significant improvement of the approximate low energy eigenstates.
The proposed technique can be directly applied to optimize existing NRG results \cite{newpaper}
and could be a beneficial improvement of impurity solvers in the context of dynamical mean-field theory, see e.g.~\cite{dmftreview}.

\textit{Acknowledgments.}
We acknowledge fruitful discussions with H. G. Evertz, A. Gendiar, K. Held, T. Pruschke, G. Sangiovanni, A. Toschi, and R. \v{Z}itko,  and financial support by the EU project QUEVADIS and the FWF SFB project ViCoM. The computational results were in part achieved using Vienna Scientific Cluster.

%
%
%
%

\clearpage

\renewcommand{\theequation}{A-\arabic{equation}}  
\setcounter{equation}{0}

\appendix

\onecolumngrid
\section{Supplementary material}
In this section we sketch the algorithm proposed in the main text of the Letter, give a proof on the validity of the measure used to quantify the accuracy of approximate eigenstates, and present some further results supporting the accuracy of the proposed method.

\vspace{1cm}

\twocolumngrid

\section{Practical Algorithm}
In Table~\ref{table:algorithm} we provide a practical algorithm for the variational optimization of the NRG states.
Let us assume we are interested in the $M$ lowest energy states of a system on sites $1,\ldots,n$ which we write as \cite{franksiam}
\begin{equation}
\ket{\psi_\alpha} = \sum_{s_j} 
\bra{\mathtt{I}} \mathbf{A}^{[1] s_1} \cdots \mathbf{A}^{[n] s_n} \ket{\mu} \bigotimes_{j=1}^{n} \ket{s_j}.
\label{eq:nrgmps}
\end{equation}
Here, the left boundary vector is a trivial vector $\underline{\mathtt{I}} \equiv (1,0,0,\ldots)$ and 
$\underline{\mu} \equiv (0,\ldots,0,1,0,\ldots)$ where the one is at the position $\mu \in \{1,\ldots,M\}$.
We will only consider the case without employing symmetries as the generalization is straight forward.
We will use a MPO representation for the Hamiltonian operator,
\[
H = \sum_{ \{s_j,s'_j\} }
\bra{\mathtt{I}} \mathbf{O}^{[1] s'_1, s_1} \cdots \mathbf{O}^{[n] s'_n, s_n}
\ket{\mathtt{I}}
\bigotimes_{i=1}^{n} \ket{s'_j}\bra{s_j}.
\]
We assume real arithmetics, the generalization to complex arithmetics is trivial.

\begin{table}
\noindent \textbf{Variational NRG algorithm}
\begin{enumerate}
\item Perform $n$ NRG steps to obtain $\{ \psi_{i}, i=1,\ldots,M\}$ \\(or just choose some random states).
\item Choose a site $j := n-1$, initialize $\underline{L}^{[0]}$ and $\underline{R}^{[n]}$.
\item Calculate/retrieve tensors $L_{l,p,l'}$ and $R^{[j]}_{r,q,r'}$.
\item Define $\mathbf{G}_q$, $\mathbf{R}_q$ and $\mathbf{X}$ and minimize
$f(\mathbf{X})$
\item $\mathbf{A}^{[j] s}_{l,r} := [\mathbf{X}]_{(l,s),r}$
\item If converged, then stop.
\item if ($j==n$), update $w_j$ and recalculate $\underline{R}^{[n]}$.
\item $j := j-1$ or $j := j+1$ and go to $3$.
\end{enumerate}
\caption{Sketch of the VNRG algorithm for the variational optimization of a NRG-MPS set of states.}
\label{table:algorithm}
\end{table}

In the algorithm, we employ the fundamental concepts of the matrix product state formalism, such as calculating the blocks in a recursive way,
\[
L^{[j]}_{r,q,r'} = \sum_{l,l',s,s',p} L^{[j-1]}_{l,p,l'} A^{[j-1] s}_{l,r} A^{[j-1] s'}_{l',r'} O^{[j-1] s',s}_{p,q}
\]
and similarly for the right block $R^{[j]}_{r,q,r'}$. The starting block is trivial, $\underline{L}^{[0]} \equiv \mathbf{I}_{1\times 1 \times 1}$.
The only step which differs from the MPS formalism is initialization of the right block $\underline{R}^{[n]} \in \mathbb{R}^{M \times 1 \times M}$ which now contains the sum over $M$ energies and can be written as 
\[
R^{[n]}_{r,q,r'} \equiv \delta_{q,0} \delta_{r,r'} w_r.
\]
where $w_r$ is some chosen weight function, e.g. $w_r = 1$ (no weighting) or $w_r = {\rm e}^{-r}$ (position weighting) or $w_r = {\rm e}^{-(E_r - E_0)}$ (energy weighting). In case of energy weighting, we use the initial approximation for the energies and re-calculate the boundary block after each sweep.

For the unitary optimization we formally reshape tensors $\underline{L}^{[j]} \cdot \underline{R}^{[j]}$,  $\underline{O}^{[j]}$, and $\underline{A}^{[j]}$ to matrices $\mathbf{G}_q$, $\mathbf{R}_q$ and $\mathbf{X}$ as $[\mathbf{G}_q]_{(l,s),(l',s')} = \sum_{p} L^{[j]}_{l,p,l'} O^{[j] s' s}_{p,q}$, 
$[\mathbf{R}_q]_{i,j} \equiv R_{i,q,j}$ and $[\mathbf{X}]_{(l,s),r} \equiv A^{[j] s}_{l,r}$.
The idea behind this matrix reshaping is that we can now use some black-box routine which minimizes the following cost function
\begin{equation}
f(\mathbf{X}) = \sum_q {\rm tr} \big(  \mathbf{G}_q^T \mathbf{X} \mathbf{R}_q \mathbf{X}^T \big)
\label{eq:cost}
\end{equation}
under the constraint that $\mathbf{X}^T \cdot \mathbf{X} = \mathbf{1}$. The algorithm which gives the best results according to our experience was proposed in Ref.~\cite{abrudanCG}.

\textit{Relation to the NRG.}
The MPS description of our description~(\ref{eq:nrgmps}) is tightly connected to the NRG algorithm where one 
diagonalizes an effective hamiltonian $H_{N+1}$ (Eq. 42 in \cite{bullareview}) and obtains 
eigenvalues $E_{N+1}$ and eigenvectors $\ket{w}_{N+1}$ which are related to the eigenvectors of the smaller system ($N$ sites) as (Eq. 43 in \cite{bullareview})
\[
\ket{w}_{N+1} = \sum_{rs} U(w,rs) \ket{r; s}_{N+1}.
\]
The coefficients $U(w,rs)$ in the above relation are identical to the tensor coefficients $A^{[N+1] s}_{r,w}$ used in~(\ref{eq:nrgmps}).

\textit{Relation to DMRG with targeting}.
The main difference between the DMRG with targeting (DMRGt) and the VNRG is that the former optimizes the set of states by moving the external bond along the chain whereas in the latter the external bond is fixed to a chosen site. In the DMRGt, the energies (and thus our cost function) do not monotonically decrease when moving the bond along the chain but there exists a site in the chain where the energies are minimal. Typically, in models with translation invariance such as the Ising model or the Heisenberg model, the highest accuracy is reached when at the center of the chain. In impurity models with falling hopping terms, however, the optimal site to put the external leg is (for long chains) the end of the chain - in which case the cheaper variant (scaling as $D^3$) of the VNRG is preferable. 
We have described the cheaper variant of the VNRG in the manuscript but we have skipped the more expensive one where the external leg is associated with some inner site. In fact, to reduce computational complexity, we do not actually associate the external leg to a site but rather introduce an additional 3-dim tensor on the bond connecting two sites: the third leg (the external one, similarly to the ``physical'' leg in the matrix product state) plays the role of the external leg. This way, the optimization of the more expensive variant of the VNRG is as follows. All tensors associated to physical sites are optimized as in the cheaper VNRG, that is minimizing the cost function~(\ref{eq:cost}) under a unitary constraint. This in total costs $n O(D^3)$ for one sweep along the chain. The additional tensor which selects the eigenstates is optimized in the same way as all tensors in the DMRG with targeting. The DMRG with targeting involves finding $M$ lowest eigenstates of a sparse $(D d)^2 \times (D d)^2$ matrix and similarly in our case where we seek $M$ lowest eigenstates of a sparse $D^2 \times D^2$ matrix. The cost of the DMRGt is thus $n O( {\tilde M} (D d)^3)$ for one sweep whereas the equivalent step contributes $O({\tilde M} D^3)$ to the VNRG, just once in a sweep.

\section{On accuracy of the eigenstates}

\textit{Statement.}
Let $H$ be a real symmetric linear operator with eigenvectors $\{ \Psi_j \}$ such that $H \ket{\Psi_j} = E_j \ket{\Psi_j}$,
and let ${\tilde \Psi_j}$ be an approximation to $\Psi_j$ with fidelity $F = \vert \braket{\Psi_j}{\tilde\Psi_j}\vert$.
If $F > \frac{1}{\sqrt{2}}$, then
\begin{equation}
1-F \leq \frac{\sqrt{2}}{\Delta_j} 
\Big( \bra{\tilde\Psi_j} H^2 \ket{\tilde\Psi_j} - \bra{\tilde\Psi_j} H \ket{\tilde\Psi_j}^2 \Big)^{1/2}
\end{equation}
where $\Delta_j = {\rm min}_{j'\neq j} \vert E_j-E_{j'}\vert$ is the spectral gap to the nearest eigenlevel. We have assumed that all considered states are normalized with respect to the Euclid norm. We assume real algebra (generalization to complex algebra is trivial).

\textit{Proof.}
Let us write the approximation ${\tilde \Psi}_j$ as 
\begin{equation}
\ket{\tilde\Psi}_j = \sqrt{1-\epsilon^2} \ket{\Psi_j} + \epsilon \ket{\Phi_j}
\end{equation}
where $\epsilon \geq 0$, $\braket{\Phi}{\Psi_j} = 0$ and $\braket{\Phi}{\Phi} = \braket{\Psi_j}{\Psi_j}=1$.
It is easy to show that the measure $G^2\equiv \bra{\tilde\Psi_j} H^2 \ket{\tilde\Psi_j} - \bra{\tilde\Psi_j} H \ket{\tilde\Psi_j}^2 $ can be written as 
\begin{equation}
G^2 = \epsilon^2 \bra{\Phi_j} (H-E_j)^2 \ket{\Phi_j} - \epsilon^4 (\bra{\Phi_j}H\ket{\Phi_j}-E_j)^2
\end{equation}
or as an expression for $\epsilon$ as
\begin{equation}
\epsilon^2 = \frac{G^2}{\bra{\Phi_j} (H-E_j)^2 \ket{\Phi_j} } + 
\epsilon^4 \frac{ (\bra{\Phi_j}H\ket{\Phi_j}-E_j)^2 }{  \bra{\Phi_j} (H-E_j)^2 \ket{\Phi_j}  }.
\end{equation}
The second term on r.h.s is upper bounded by $\epsilon^4$,
\begin{equation}
\epsilon^2 (1-\epsilon^2) \leq \frac{G^2}{\bra{\Phi_j} (H-E_j)^2 \ket{\Phi_j} } .
\end{equation}
The denominator is lower bounded by $\Delta_j^2$ which,
assuming $\epsilon^2 < \frac{1}{2}$ and thus $(1-\epsilon^2)^{-1} < 2$, finally gives us the upper bound on $\epsilon$
and the lower bound on fidelity $F = \sqrt{1-\epsilon^2} \geq 1-\epsilon$  (for $\epsilon\leq 1$) as
\begin{equation}
\epsilon^2 < 2 \frac{G^2 }{\Delta_j^2 }
\quad\Longrightarrow\quad F \geq 1 - \sqrt{2} \frac{G}{\Delta_j}. \quad \blacksquare
\end{equation}
The only assumption we made was $\epsilon^2 < \frac{1}{2}$ and $\Delta_j \neq 0$.

\begin{figure}
\centering
\includegraphics[width=\columnwidth]{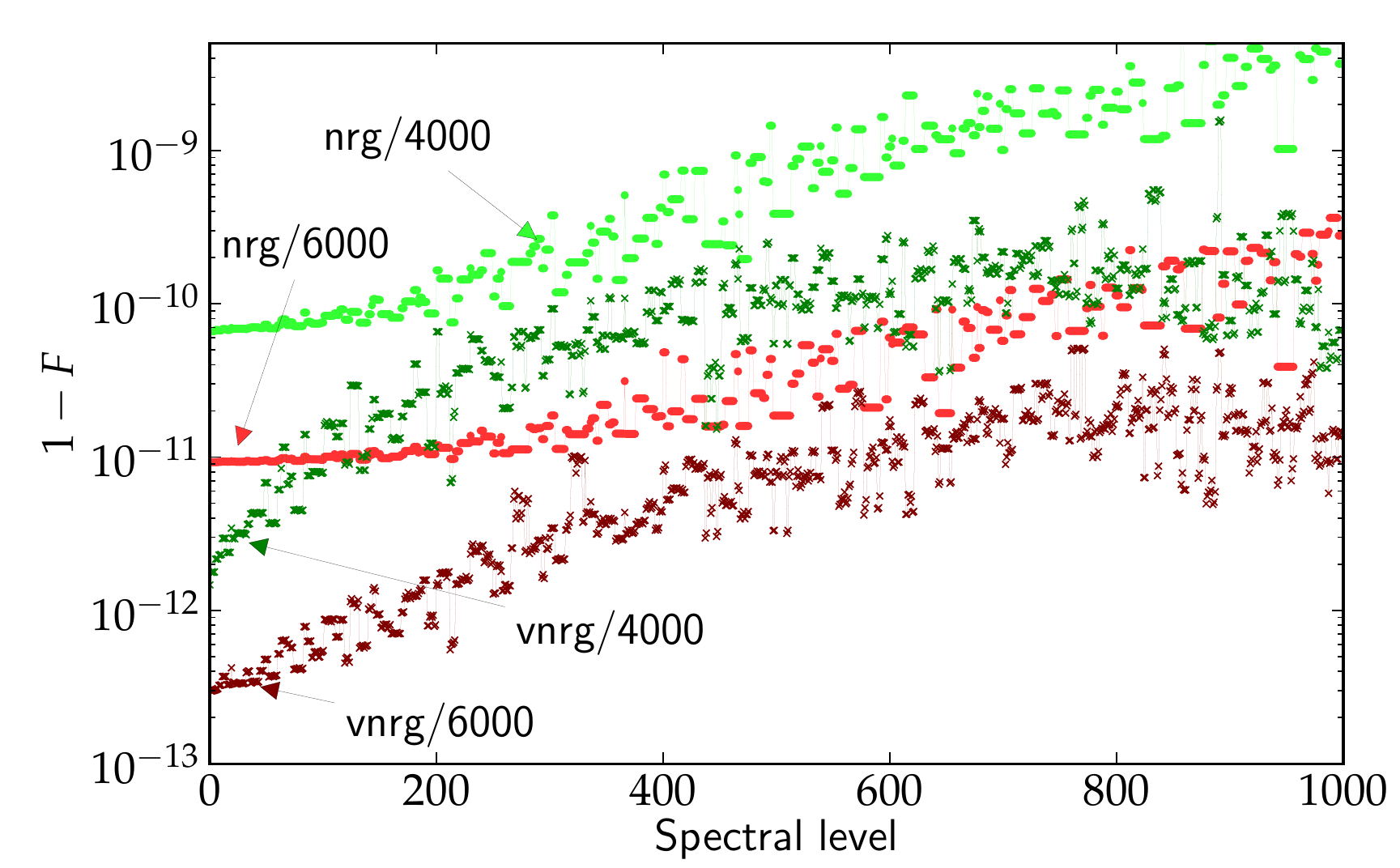}
\caption{Fidelity of the first $1000$ approximate eigenstates of the SIAM on $40$ ($N=38$)  sites obtained by NRG and vNRG and total number of kept states $M \in \{4000,6000\}$.
We have set $U=-2 \epsilon_f = 0.1$, $\xi_0=0.01$, and $\Lambda=2$ and used (sub-sector) bond dimension $D=5000$.  The basis for comparison are the NRG results with $M=16000$ kept states (and $D=6000$).}
\label{fig:fidelity}
\end{figure}

On Figure~\ref{fig:fidelity} we present the fidelity for the interacting single impurity Anderson model (SIAM) with respect to very precise results obtained with the total number of states $M=16000$ and a bond dimension $D=6000$ in individual subsectors, such that the actual number of states in individual subsectors after truncation to in total $M$ states is always smaller than $D$ and thus no additional truncation is made in subsectors. We compare the results obtained with $M=4000$ and $M=6000$, in both cases again using a sufficiently large $D$.
Let us stress that we compare the NRG and vNRG states to the very precise states obtained by the NRG which were not further optimized in order to avoid possibly unfair comparison with preference for vNRG (no change in the fidelity is observed even if we do optimize the results for $M=16000$).
We observe that the accuracy of the approximate eigenstates is always improved by the variational method proposed in the manuscript. The improvement is most significant for those states with a low energy, since the higher excited states are more entangled%
(see e.g. V. Alba, M. Fagotti, P. Calabrese, J.Stat.Mech.0910, P10020, \textbf{2009} for an analysis on integrable spin systems)
and require a larger bond dimension in the description with matrix product states.

\begin{figure}
\includegraphics[width=\columnwidth]{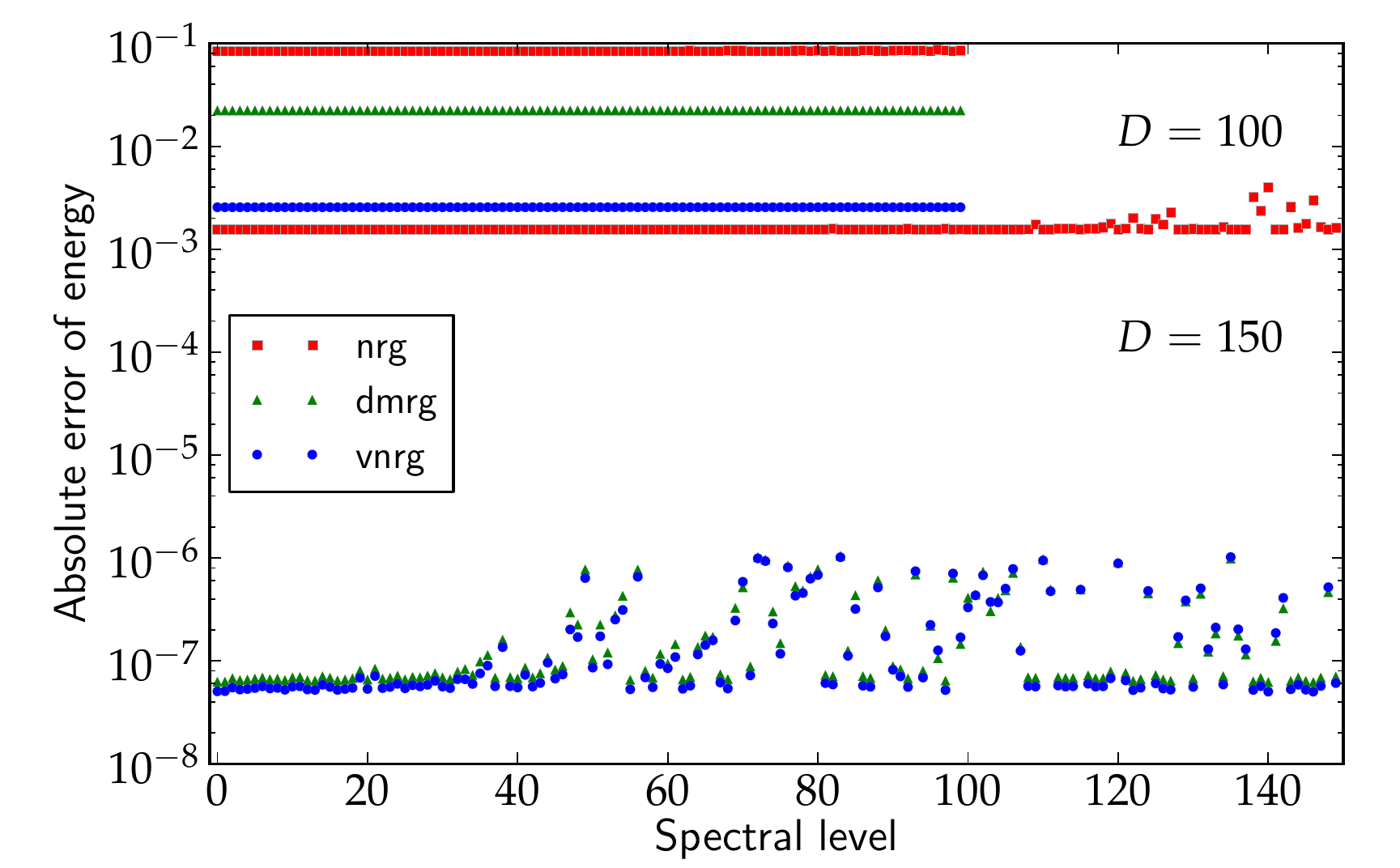}\\
\includegraphics[width=\columnwidth]{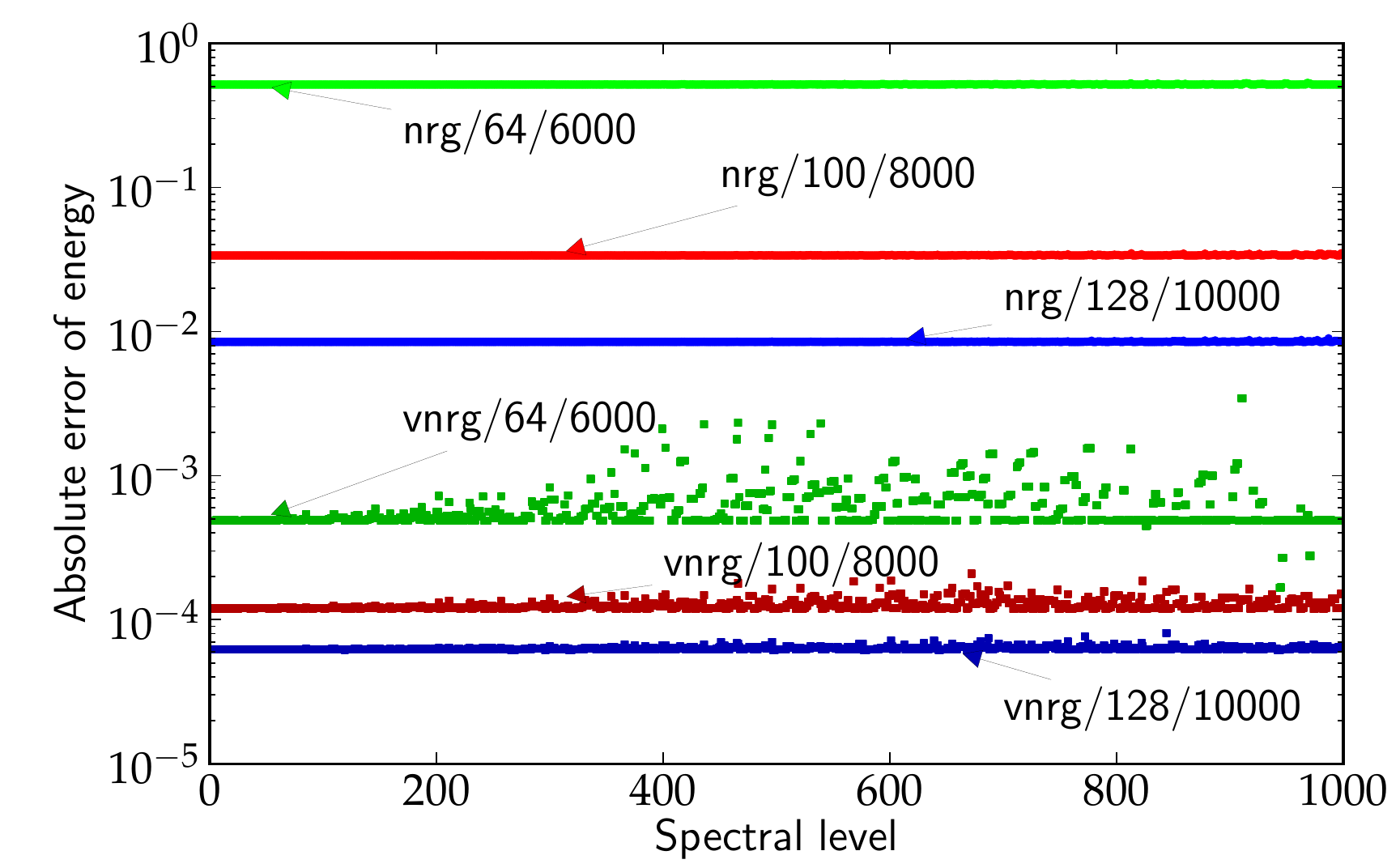}\\
\caption{Absolute difference of energies of the non-interacting SIAM (top) and the interacting SIAM (bottom). Compare to Figs. 3 and 4 in the Letter.}
\label{fig:energydiff}
\end{figure}
A single approximate eigenvalue in general does not tell anything about the quality of the corresponding approximate eigenvector except for the ground state (or the state on the opposite side of the spectrum). We consider the first $M$ lowest lying states, so the situation might be similar also in our case but we are not able to prove it.
On the other hand, good eigenvectors automatically also mean good eigenvalues, seen from the fact that the eigenvalue error is bounded from above by 
\begin{equation}
(\bra{\tilde\Psi} H \ket{\tilde\Psi} - E )^2 \leq \epsilon^4 \Delta_j^2.
\end{equation}
We observe that the results for the difference in energies, $\vert \langle H \rangle - E\vert$, and the measure $\langle H^2 \rangle - \langle H \rangle^2$ are qualitatively very similar, as shown on Fig.~\ref{fig:energydiff} where we plot the error of energies for the same data used as in Figs. 3 and 4 in the Letter.

\section{Some additional results}

\begin{figure}
\includegraphics[width=\columnwidth]{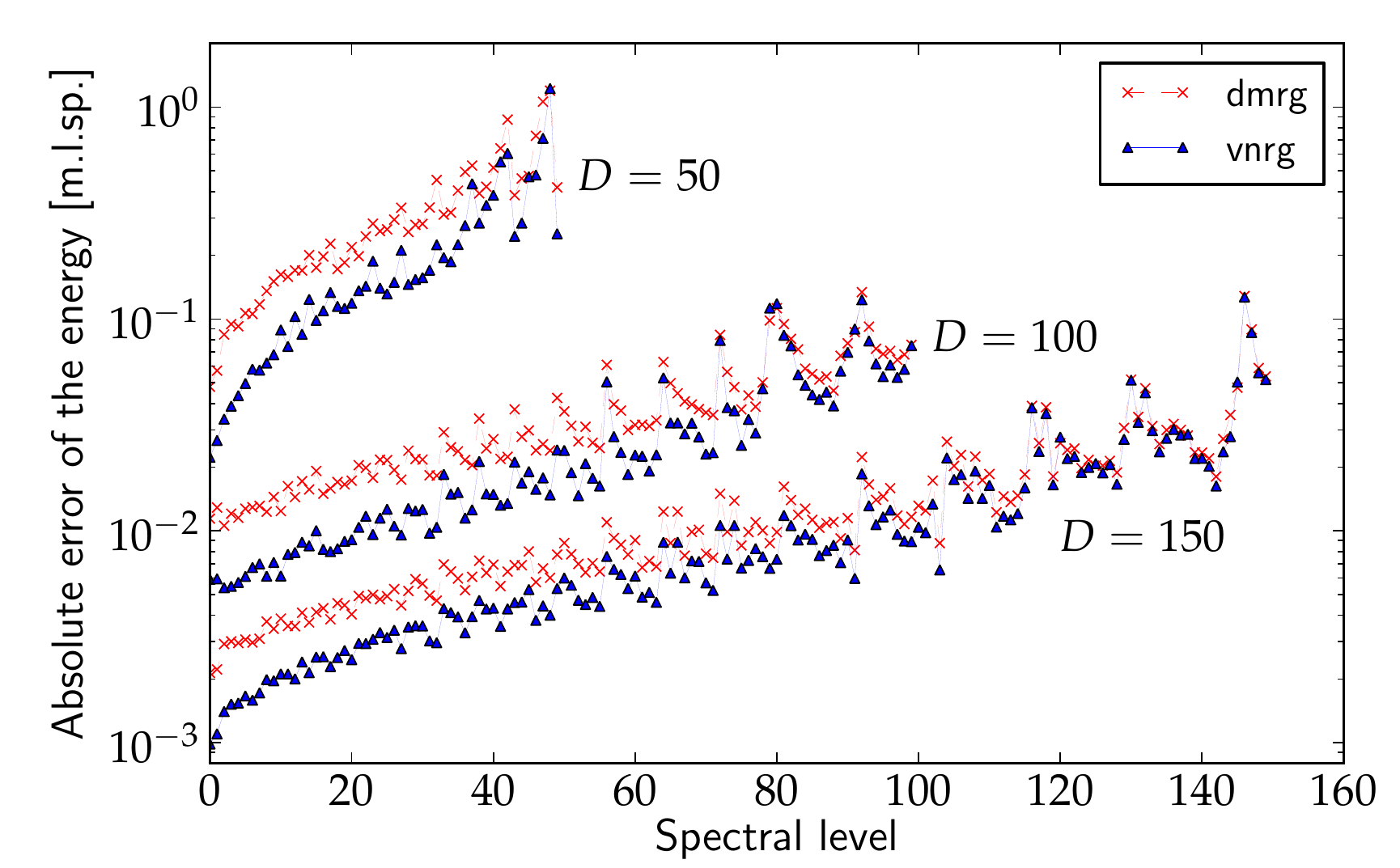}\\
\includegraphics[width=\columnwidth]{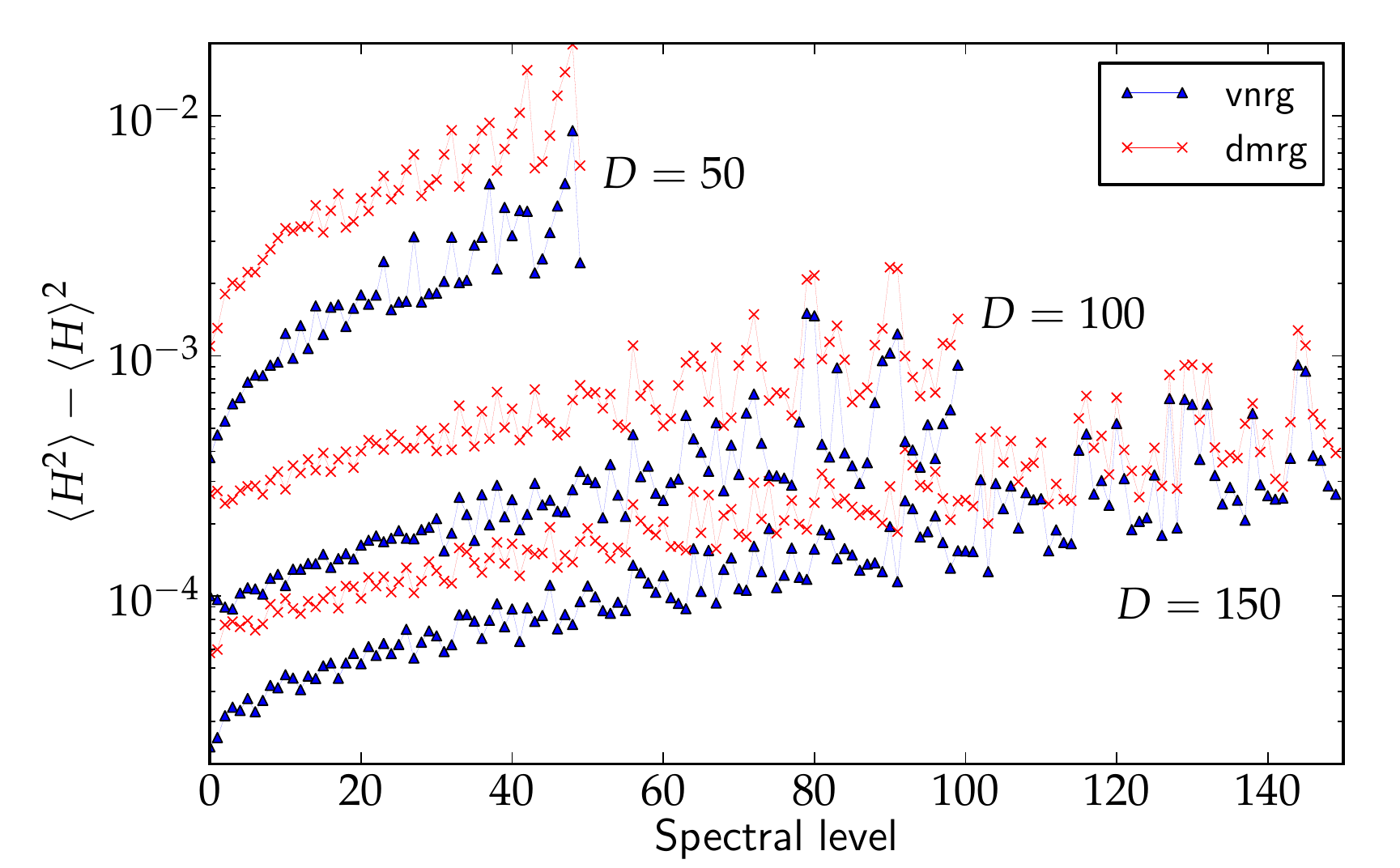}
\caption{Absolute error of energies (top) and the measure $\langle H^2 \rangle - \langle H \rangle^2$ for the quantum Ising chain of $200$ sites in critical magnetic field.}
\label{fig:criticalising}
\end{figure}

We also test the variational NRG method on an example of quantum Ising chain in critical transverse magnetic field. Similar to the tilted Ising presented in the Letter, we first calculate $D$ excited states with a bond dimension $D$ by means of the DMRG with targeting and then improve the results using the variational scheme. The quantum Ising chain in transverse field is a quadratic model and as such exactly solvable which allows us to compare the obtained energies (eigenvalues) to the exact values. We observe (Fig.~\ref{fig:criticalising}) a very good qualitative agreement between the measure $\langle H^2\rangle - \langle H \rangle^2$ which quantifies the accuracy of the eigenstates, and the absolute errors of the energies. Note that in this case, the improvement on the DMRG is more significant, since the model is critical and the excited states carry more entanglement - and thus demand a higher bond dimension in the MPS description.

 \section{Expectation values}
The accuracy of the expectation values is connected to the fidelity. The zero temperature Green 
function is determined by matrix elements $\bra{\Psi_j} d_{\sigma} \ket{0}$ and the corresponding energy differences $E_j - E_0$ where $E_0$ is the energy of the ground state, and also 
$\bra{\Psi_j} d_{\sigma}^\dagger \ket{0}$ and the corresponding energy differences, see \cite{bullareview} for details. To obtain the complete Green function, one would have to merge contributions from all NRG patches (i.e. for all system sizes) which is beyond the scope of this manuscript. We instead only calculate the ten largest matrix elements mentioned before and compare the accuracy before and after the variational optimization and show the results in Table~\ref{fig:table1}. We observe that by the variational optimization we gain roughly one digit of accuracy in the matrix elements.

\begingroup \squeezetable

\begin{table}
\centering

\begin{tabular}{||r|rr|rr||}
\hline
\hline
vnrg/10000 & nrg/4000 & vnrg/4000 & nrg/6000 & vnrg/6000 \\
\hline
0.0122411109 & 0.012240{\color{red}9812}   & 0.0122411{\color{red}285}   & 0.012241{\color{red}0278}   & 0.0122411{\color{red}168}  \\
\hline
0.0115338997 & 0.011533{\color{red}8108}   & 0.01153390{\color{red}45}   & 0.011533{\color{red}8277}   & 0.0115339{\color{red}101}  \\
\hline
0.0106204130 & 0.010620{\color{red}3467}   & 0.010620{\color{red}4772}   & 0.010620{\color{red}3414}   & 0.0106204{\color{red}360}  \\
\hline
0.0088452899 & 0.008845{\color{red}2096}   & 0.0088453{\color{red}015}   & 0.008845{\color{red}2296}   & 0.00884529{\color{red}34}  \\
\hline
0.0071215907 & 0.007121{\color{red}5287}   & 0.0071215{\color{red}980}   & 0.0071215{\color{red}428}   & 0.0071215907 \\
\hline
0.0066356189 & 0.0066{\color{red}202112}   & 0.0066{\color{red}266177}   & 0.006635{\color{red}3419}   & 0.006635{\color{red}3341}  \\
\hline
0.0051098361 & 0.0051098{\color{red}100}   & 0.0051098{\color{red}444}   & 0.0051097{\color{red}957}   & 0.0051098{\color{red}185}  \\
\hline
0.0049692462 & 0.0049{\color{red}637898}   & 0.00497{\color{red}21638}   & 0.004969{\color{red}3508}   & 0.0049692{\color{red}198}  \\
\hline
0.0039662029 & 0.0039661{\color{red}700}   & 0.0039662{\color{red}269}   & 0.0039661{\color{red}751}   & 0.0039662{\color{red}175}  \\
\hline
0.0034311308 & 0.003{\color{red}3144374}   & 0.003{\color{red}3174038}   & 0.003430{\color{red}7370}   & 0.003430{\color{red}6428}  \\
\hline
\hline
\end{tabular}

\caption{The largest ten values of $\vert \bra{\psi_j} d_\sigma \ket{0}\vert$ for the 
SIAM on $40$ sites,
$U=0.1$, $\Lambda=2$, $\xi_0=0.01$, $\epsilon_f=-0.05$.
The unreliable digits are colored red.}
\label{fig:table1}
\end{table}
\endgroup

\begin{figure}
\includegraphics[width=\columnwidth]{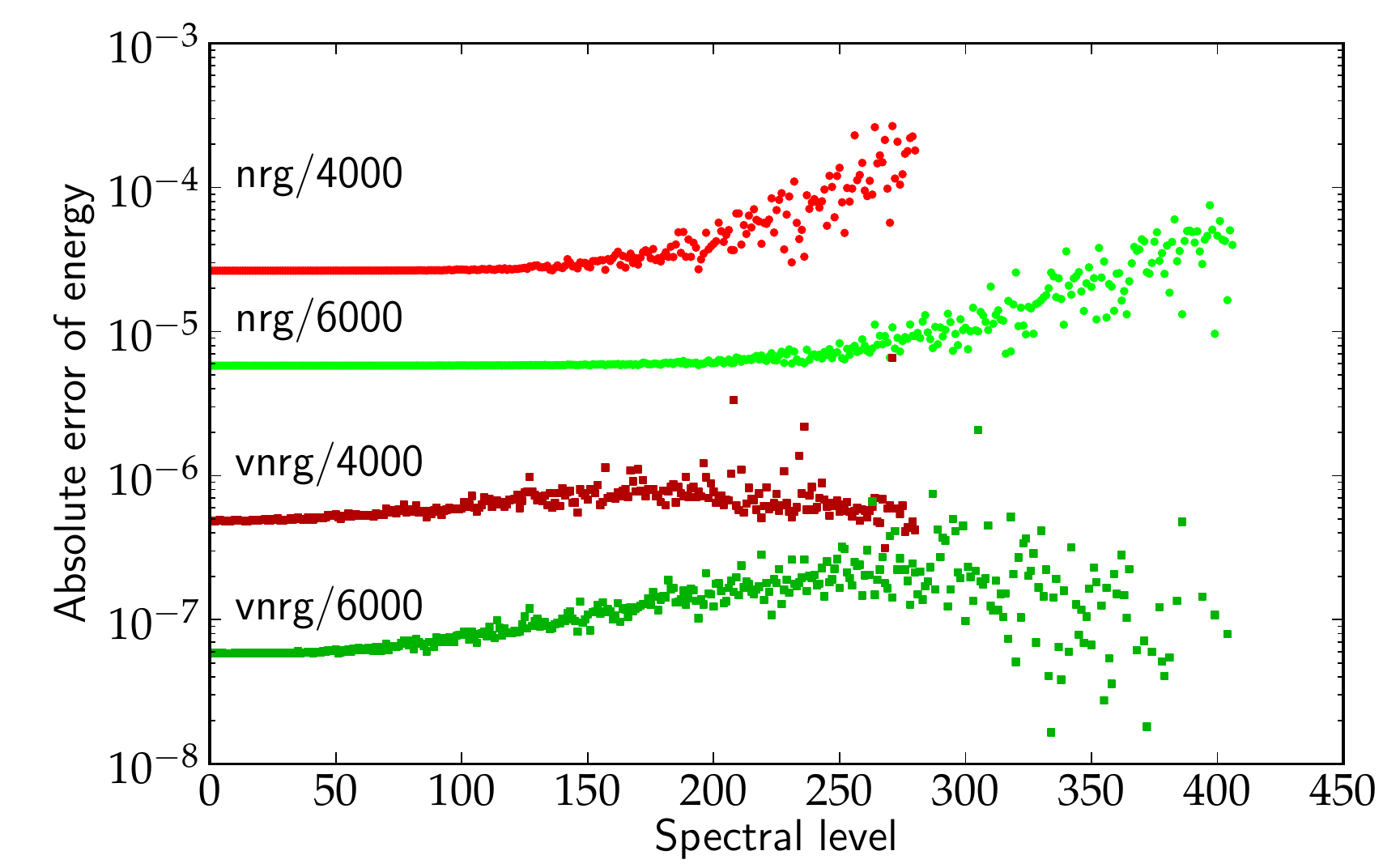}
\caption{Absolute error of energies in the subsector $d_{\sigma} \ket{0}$ for SIAM on $40$ sites, $U=0.1$, $\Lambda=2$, $\xi_0=0.01$, $\epsilon_f=-0.05$.}
\label{fig:subsector}
\end{figure}

As seen from the matrix elements, the only states that contribute to the Green functions are those which have one particles less or more than the ground state, that is the states which live in the 
subsectors $d_{\sigma} \ket{0}$ and $d_\sigma^\dagger \ket{0}$, respectively. We may therefore restrict the optimization to only these two sector which enhances the accuracy and reduces the computational costs. The ground state can be optimized independently from that by the DMRG or MPS techniques. In Fig.~\ref{fig:subsector} we show that improvement of the energies in the subsector containing the state $d_\sigma \ket{0}$ where the basis of comparison are the NRG results with $M=16000$. In this case we use sufficiently large sub-sector bond dimension $D=6000$ such that no additional truncation is done in the sub-sectors and we take the NRG results where in total $M \in \{4000,6000\}$ states were kept.

For finite-temperature Green functions we also need excited states in other symmetry subsectors with respect to the total particle number and total $S^z$. However, they can all be optimized independently and in parallel.

\end{document}